\documentstyle[aps,prl,twocolumn]{revtex}
\draft
\def\p{\scriptstyle +1}
\def\m{\scriptstyle -1}

\begin{document}

\title{\null\vspace*{-1truecm}  \hfill\mbox{\small IISc-CTS-1/00}\\
            \vspace*{-0.2truecm}\hfill\mbox{\tt\small quant-ph/0012149}\\
\vspace*{0.5truecm}Quantum Database Search can do without Sorting}
\author{Apoorva Patel}
\address{CTS and SERC, Indian Institute of Science, Bangalore-560012\\
{\small E-mail: adpatel@cts.iisc.ernet.in}}
\date{{31 May 2001}}
\maketitle

\begin{abstract}
\noindent
Sorting is a fundamental computational process, which facilitates subsequent
searching of a database. It can be thought of as factorisation of the search
process.  The location of a desired item in a sorted database can be found
by classical queries that inspect one letter of the label at a time. For an
unsorted database, no such classical quick search algorithm is available.
If the database permits quantum queries, however, then mere digitisation is
sufficient for efficient search. Sorting becomes redundant with the quantum
superposition of states. A quantum algorithm is written down which locates
the desired item in an unsorted database a factor of two faster than the
best classical algorithm can in a sorted database. This algorithm has close
resemblance to the assembly process in DNA replication.
\end{abstract}
\pacs{PACS: 03.67.Lx, 87.15.By}

Consider a collection of items in a database, characterised by some property
with values on the real line. Sorting these items means arranging them in an
ordered sequence according to their property values. Though property values
along the real line are sufficient for establishing the sequence, the values
are taken to be distinct, and without loss of generality they can be replaced
by integer labels. The integer labels can be easily digitised, i.e. written
as a string of letters belonging to a finite alphabet. When the alphabet has
$a$ letters, a string of $n$ letters can label $N=a^n$ items. (If the number
of items in the database is not a power of $a$, then the database is padded
up with extra labels to make $N=a^n$.) In digital computers, this finite
alphabet has the smallest size, i.e. $a=2$, and the letters are called bits.

The purpose of sorting is to facilitate subsequent searches of that database
for items with known values of the property. Though digitisation does not
change the order of the items in the sequence, it simplifies the search
process. One does not look for the complete string of letters in one go,
but sequentially looks for one letter at a time. A dictionary is a simple
example of a sorted database with alphabetical ordering. With digitisation,
the individual search steps have to distinguish amongst only a limited number
of possibilities, and the maximum simplification of the search steps occurs
when the alphabet has the smallest size, i.e. $a=2$. For a database of size
$N$, sorting requires $O(N \log_2 N)$ operations \cite{knuth}.
Once a database is sorted, all subsequent searches to locate any item in
it require only $O(\log_2 N)$ queries. Search for a desired item in an
unsorted database would take $O(N/2)$ queries, provided that the search
process has a memory so that an item rejected once is not picked up again
for inspection. This reduction in search effort is what makes the laborious
process of sorting worthwhile, to be carried out once and for all.

The speed-up achieved by sorting can be understood as factorisation of
the search process. An alphabet with a finite number of letters can be
looked upon as a group of finite order, and any function defined on such
a group is expressible as a polynomial. A function defined with multiple
letters becomes a multivariable polynomial. A single variable polynomial
can always be fully factored over the domain of complex numbers, but a
multivariable polynomial may not be. When a polynomial can be factored,
it is much more efficient to evaluate it by multiplying all the factors
together than by calculating it term by term. The search function is such
that it can be fully factored, resulting in a speed-up of $O(N/\log_2 N)$.
As a simple case consider a database of $N=2^n$ items. Suppose the desired
item has the label
\begin{equation}
x \equiv x_1x_2x_3x_4 \ldots x_n = 1011 \ldots 0 ~.
\end{equation}
Any function of $x$ reduces to a polynomial in $x_i$ because $x_i\in\{0,1\}$
implies $x_i^2=x_i$. The search process is equivalent to finding $x$ such that
\begin{equation}
f(x) = \prod_i f_i(x_i) = x_1(1-x_2)x_3x_4 \ldots (1-x_n)
\end{equation}
equals $1$. $f(x)$ evaluates to zero for all the other items in the database.
In the sorted and digitised database, one searches for an item by inspecting
only one $x_i$ at a time, i.e. evaluating $f(x)$ by sequentially combining
its factors $f_i(x_i)$.

In the language of algorithms, the function $f(x)$ is referred to as an
oracle. The capability of evaluating $f(x)$ for any $x$ is assigned to a
black box, and the efficiency of algorithms is judged by the number of times
the oracle is queried. To take advantage of a sorted database, it is crucial
to have the factorised oracles, $f_i(x_i)$, available. If only the oracle
$f(x)$ is available, then the search process requires $O(N/2)$ queries even
with a sorted database. The collection of factorised oracles $f_i(x_i)$ is
more powerful than the global oracle $f(x)$---one can construct $f(x)$ by
combining all the $f_i(x_i)$ together, but the individual $f_i(x_i)$ cannot
be determined given only $f(x)$. Generally, how an oracle is physically
implemented is not of concern in the complexity analysis of an algorithm.
But it is clear that $f_i(x_i)$ are easier to physically implement than
$f(x)$, because $x_i$ span a much smaller range of values than $x$.

With a single query, the factorised oracles are able to characterise a
group of items (e.g. all the items with $1$ as the first bit of the label),
unlike the global oracle that can characterise only a single item.
With a sorted database, one takes advantage of this property of factorised
oracles. The items that are addressed together by factorised oracles are
arranged together in a sorted database, and not left scattered all over the
place. The factorised oracles then easily pick up only the relevant group
of items, and implement the search process one letter at a time, whereby
the size of the group reduces in a geometric progression. If the items are
not grouped together and left scattered all over, the factorised oracles do
not offer any advantage compared to the global oracle.

Classically, the only way to group items together is to arrange them close
to each other, as in a sorted database. Quantum physics provides another
mechanism for grouping items together---quantum superposition of states.
Superposed states are grouped together in the sense that they all correspond
to the same physical location (in space as well as in time), and they all
can be simultaneously acted upon by a single physical transformation.
In a quantum search process, the inspection of items need not be carried
out picking one item at a time from the database; instead the search oracles
can be applied directly to groups of items superposed together. In what
follows, I demonstrate that when factorised oracles are available, quantum
superposition can be used to make sorting redundant.

The quantum parallelism strategy, based on superposition of states, is quite
general. It makes the algorithms more efficient, and it has been exploited
to convert classically superpolynomial problems into quantum polynomial ones.
Because of linearity of quantum mechanics, when the initial state is a quantum
superposition of states, the final state after carrying out the algorithmic
operations will be a quantum superposition of the corresponding outcomes.
Of course, at the end, quantum measurement does not permit individual
determination of each outcome. But if only a single property of the possible
outcomes is desired, then it may be possible to pick it up by a cleverly
designed measurement (perhaps involving interference of the possible
outcomes). The net result is a substantial speed-up of the algorithm.

Quantum algorithms need to obey an important practical constraint, i.e.
all their communications with the outside world must have classical
representations. These communications include input, output and oracles.
This constraint arises from our own limitations as classical observers.
Our problems are defined in a classical language, and we do not know how
to make sense out of them otherwise. So we always have to find ways to
extract classical meanings from quantum superpositions.

To take maximal advantage of quantum parallelism, many quantum algorithms
start with a uniform quantum superposition of all possible classical input
states. Such a state can be easily created from a single classical state
by a fully factorised operation. For instance, each qubit in an $n-$qubit
string can be put into an equal superposition of its two eigenstates, by
starting with a zero-string and applying the Walsh-Hadamard transformation:
\begin{equation}
H = {1\over\sqrt{2}}\left( \matrix{1&1 \cr 1&-1 \cr} \right) ~,~~
H |0\rangle = {1\over\sqrt{2}} (|0\rangle+|1\rangle) ~,
\end{equation}
\begin{equation}
|00 \ldots 0\rangle \longrightarrow \prod_{i=1}^n (H|0\rangle)_i
  = \sum_{j=1}^N {1\over\sqrt{N}} |j\rangle ~.
\end{equation}
Digitisation of the state labels is implicit but essential in this process. 
It reduces the spatial resources needed to solve the problem, i.e. $2^n$
states are encoded using only $n$ qubits.
An obvious feature of superposition is that it is commutative---the order
in which the states are superposed is irrelevant. This is the property
that makes sorting redundant whenever the states are combined in a
superposition.

Grover's quantum database search algorithm \cite{grover}
starts with a maximally superposed state, uses a binary search function
and reaches a single state at the end. Thus the input, the output and the
oracle have classical representations as desired. Grover's algorithm is
directly applied to an unsorted database. The unitary quantum operators
used are the search function $F$ (which evaluates to $-1$ when the labels
match and to $+1$ when they do not) and an operation $R$ that reflects all
the amplitudes about their average. Alternate application of these two
operators takes the maximally superposed state to the desired state.
\begin{equation}
|x\rangle = \underbrace{(R F) \ldots\ldots (R F)}_{O(\sqrt{N}) \rm times}
	    ~(\sum_{j=1}^N {1\over\sqrt{N}} |j\rangle) ~.
\end{equation}
Compared to classical unsorted database search, quantum parallelism provides
a quadratic speed-up and the desired item is located with $O(\sqrt{N})$
queries. This rate, however, is much slower than that for the classical
sorted database search, and we can look for ways to boost it.

Though sorting has become redundant, Grover's algorithm can still be
factorised. Factorisation of the search function requires digitisation
of the database, so let us start with a digitised database. (There is no
hindrance to digitising a database, even if it is unsorted. All that is
required is that the items be distinct from each other.) Digitisation
makes it possible to look at one letter of the label at a time. The search
process will inspect each letter of the label in turn, and decide whether
it matches the corresponding letter of the desired string or not.
Since Grover's algorithm requires only one query to pick one item out of
four with certainty, it is efficient to digitise the quantum database
with $a=4$. At the $i^{\rm th}$ step of the algorithm, let $F_i$ be the
single-letter search function which looks at the $i^{\rm th}$ letter of
the label. $F_i$ will evaluate to $-1$ when the letters match and to $+1$
when they do not.
\begin{equation}
F_i = I_1 \otimes \ldots \otimes \{\pm1\}_i \otimes \ldots \otimes I_n ~.
\end{equation}
The reflection operation $R_i$ that goes along with $F_i$ can also be
written in a factorised form.
\begin{equation}
R_i = I_1 \otimes \ldots \otimes
{\scriptstyle{1\over2}} \left( \matrix{\m&\p&\p&\p \cr
                                       \p&\m&\p&\p \cr
                                       \p&\p&\m&\p \cr
                                       \p&\p&\p&\m \cr} \right)_i
\otimes \ldots \otimes I_n ~.
\end{equation}
When $R_i$ follows $F_i$, it carries out the same task as $R$, i.e. reflects
all the amplitudes about their average. Since $4$ is a power of $2$, the
non-trivial transformation matrix in $R_i$ can be decomposed in qubit
notation as $(H \otimes H) F_0 (H \otimes H)$, where $F_0$ and $H$ follow a
common qubit basis convention and $F_0$ flips the sign of the state labeled
$0$. It is not mandatory, however, that $R_i$ be implemented in this manner.

In addition, we require the projection/measurement operators $P_i$,
for every letter of the label. Once the $i^{\rm th}$ letter gets fixed
by $P_i$, it cannot change during the subsequent steps of the algorithm,
i.e.  $P_i$ eliminates from the quantum Hilbert space all the states whose
$i^{\rm th}$ letter does not match with the desired label. Then the
factorised quantum search algorithm is:
\begin{equation}
|x\rangle = \prod_{i=1}^n (P_i R_i F_i)
	    ~(\sum_{j=1}^N {1\over\sqrt{N}} |j\rangle) ~.
\end{equation}
The starting point is the uniform superposition of all the $N$ states.
Application of $R_1 F_1$ doubles the amplitudes of all those strings whose
first letter matches with the first letter of the desired item label, and
reduces all the other amplitudes to zero. The projection/measurement
operator $P_1$ does not change any of the amplitudes, but removes from the
Hilbert space all the states with zero amplitude. The intermediate result
is then a uniform superposition over the reduced Hilbert space of dimension
$N/4$, containing all the states whose first letter matches with the desired
item label. One by one, the algorithm goes through all the letters of the
string and ultimately finds the item with the desired label.

In the classical search of a sorted database, the sequence of $i$ in which
the oracles have to be queried is determined by the convention used for
grouping the items together during the sorting process. For example, with
the conventional labeling as in a dictionary, one has to first query the
first letter using $f_1(x_1)$, then query the second letter using $f_2(x_2)$,
and so on. For the factorised quantum algorithm, Eq.(8) above, there is no
such requirement; the database is unsorted and the sequence of $i$ during
the execution of the algorithm does not matter.

The use of $P_i$ is essential here, so that the eliminated states do not
take part in the subsequent steps of the algorithm, i.e. $R_i$ acts only
on the Hilbert space of dimension $4^{1-i}N$. In a general quantum setting,
the reduction of dimension of Hilbert space using $P_i$ is a complicated
operation. But the structure of the factorised search algorithm is such that
all the states eliminated by $P_i$ have zero amplitudes. For this limited
purpose, $P_i$ can be denoted using identity operators,
\begin{equation}
P_i = I_1 \otimes \ldots \otimes (I_i)_{\rm fix} \otimes \ldots \otimes I_n ~.
\end{equation}
Introduction of $P_i$ also provides an error correcting mechanism. If the
measured letter of the label does not match with the desired letter, due to
imperfect implementation of $F_i$ and $R_i$ or due to effects of decoherence,
then the algorithm can be restarted at the $i^{\rm th}$ step, with a uniform
superposition of $4^{1-i}N$ states. (Such a state can be prepared by setting
$(i-1)$ letters to their correctly measured values, the remaining $(n-i+1)$
letters to zero, and then applying Walsh-Hadamard transformation to these
$(n-i+1)$ letters.)

The transformations $F_i$, $R_i$ and $P_i$ act on a single letter of the
label only. The equations above show that the non-trivial operations are
performed only on the $i^{\rm th}$ letter, while the rest of the letters
remain unchanged (identity operations). Clearly the resources required to
implement such operations depend only on the size of the alphabet $a$, and
are independent of the length of the label $n$. The physical resources
needed to implement the factorised quantum search algorithm---number of
letters, number of queries and number of algorithmic steps---are therefore
all proportional to $n$.

We have thus arrived at the first important result of this article: The
factorised quantum search algorithm locates the desired item in an unsorted
database using $O(\log_4 N)$ queries, which is a factor of two improvement
over the best search algorithm for a classical sorted database. (If the
algorithm has to be restarted to correct errors, as mentioned above, then
the improvement factor will be smaller.)

The frequent intervention of projection/measurement operators in this
factorised algorithm makes it look like a quantum process that is
stabilised by intermediate classical states. Indeed, we can think of
a process closely related to search---assembling a desired string by
putting together a sequence of letters. The letters have to be available
in a jumbled up database and an oracle has to exist dictating the order
in which the letters are to be put together. The algorithm will then
carry out the assembly line operation by adding one letter at a time.
At each step, one out of $a$ items is selected, and the full string is
constructed by putting together $n$ items. The quantum assembly algorithm
requires a superposition of only $a$ states at each step, and not $N$
states as in Grover's algorithm; this reduction in the number of superposed
states should make it easier to implement in reality. The physical resources
necessary to implement the quantum assembly algorithm are again proportional
to $n$, and in terms of the number of queries made, the quantum assembly
algorithm provides a factor of two speed-up compared to the best classical
assembly algorithm.

It is too tempting to overlook a fundamental biological process that works
in this manner, i.e. replication of DNA. (As a matter of fact, biochemistry
is full of assembly processes which synthesise desired objects out of their
components by pattern recognition oracles.) The DNA alphabet has four
letters, i.e. the bases A,T,C,G. In DNA itself these bases are linked
together as a string, but individual bases also randomly float around in
the cellular environment. During replication the two strands of the DNA
double helix separate. Each intact strand duplicates itself in an assembly
line operation, picking up one base at a time from the cellular environment.
The intact strand is the factorised oracle, with the complementarity of the
bases acting as the search function. The fact that this process takes place
at the molecular scale and uses an alphabet of four letters raises a highly
provocative thought. Could it be that the evolution of life sensed the
advantage of a quantum algorithm, and opted to organise the genetic
information in DNA using four bases? Note that classically just two bases
(one complementary pair) are sufficient to carry the genetic information.
This question has been discussed elsewhere \cite{patel}.

Finally, it is worth noting that even though factorisation is a classical
computational procedure, when combined with quantum parallelism, it has
proved to be useful in constructing efficient quantum algorithms. Under the
best circumstances, factorisation can provide a speed-up of $O(N/\log_2 N)$,
and so can quantum parallelism. Shor's efficient quantum solution to the
period finding problem \cite{shor}
combines factorisation (FFT is the factorised version of Fourier transform)
and quantum parallelism. Each provides a gain of $O(N/\log_2 N)$, and the
final algorithm requires $O((\log_2 N)^2)$ steps. In the search algorithm,
gains of factorisation (i.e. sorting) and quantum parallelism overlap.
So one cannot obtain a speed-up of $O(N/\log_2 N)$ twice, and the quantum
factorised algorithm provides only a factor of two speed-up over the classical
factorised algorithm. Yet another problem, that of finding parity of a set
of bits, cannot gain anything from factorisation and very little from quantum
parallelism; the best quantum algorithm does not improve the classical one
by much (both are $O(N)$) \cite{parity}.

\section*{Acknowledgements}

This work was supported in part by the Rajiv Gandhi Institute of
Contemporary Studies in cooperation with the Jawaharlal Nehru Centre
for Advanced Scientific Research, Bangalore.
I am grateful to Lov Grover and the Optical Physics Research group of
Lucent Technologies for their hospitality during part of this work.

\end{document}